\documentclass{svmult}
\usepackage[english]{babel}
\usepackage{cite}
\usepackage{amssymb,amsmath}
\usepackage{graphicx,subfigure}
\usepackage{float}
\usepackage[bottom]{footmisc}
\usepackage{lscape}
\usepackage[basic,box]{circ}

\newcommand{\pref}[1]{(\ref{#1})}
\newcommand{\Sec}[1]{Sec.~\ref{#1}}
\newcommand{\Eq}[1]{Eq.~(\ref{#1})}
\newcommand{\Fig}[1]{Fig.~\ref{#1}}

\newcommand{\TO}{\mathbb{T}}
\newcommand{\erfc}{\mathrm{erfc}}

\begin{document}

\title*{On the electrodynamics of \\
            neural networks}
\titlerunning{Electrodynamics of neural networks}
\author{Peter beim Graben
    \and
    Serafim Rodrigues}

\institute{Peter beim Graben \\
    Bernstein Center for Computational Neuroscience Berlin, \\
    Department of German Studies and Linguistics, \\
    Humboldt-Universit\"at zu Berlin, Germany \\
    \and
    Serafim Rodrigues \\
    Centre for Robotics and Neural Systems, \\
    School of Computing and Mathematics, \\
    University of Plymouth, United Kingdom
}

\maketitle

\begin{abstract}
~
We present a microscopic approach for the coupling of cortical activity, as resulting from proper dipole currents of pyramidal neurons, to the electromagnetic field in extracellular fluid in presence of diffusion and Ohmic conduction. Starting from a full-fledged three-compartment model of a single pyramidal neuron, including shunting and dendritic propagation, we derive an observation model for dendritic dipole currents in extracellular space and thereby for the dendritic field potential that contributes to the local field potential of a neural population. Under reasonable simplifications, we then derive a leaky integrate-and-fire model for the dynamics of a neural network, which facilitates comparison with existing neural network and observation models. In particular, we compare our results with a related model by means of numerical simulations. Performing a continuum limit, neural activity becomes represented by a neural field equation, while an observation model for electric field potentials is obtained from the interaction of cortical dipole currents with charge density in non-resistive extracellular space as described by the Nernst-Planck equation. Our work consistently satisfies the widespread dipole assumption discussed in the neuroscientific literature.
\end{abstract}



\section{Introduction}
\label{sec:GS:intro}

Hans Berger's 1924 discovery of the human \emph{electroencephalogram} \cite{Berger29} lead to a tremendous research enterprise in clinical, cognitive and computational neurosciences \cite{SchomerSilva11}. However, one of the yet unresolved problems in the biophysics of neural systems is understanding the proper coupling of complex neural network dynamics to the electromagnetic field, that is macroscopically measurable by means of neural mass potentials, such as \emph{local field potential} (LFP) or electroencephalogram (EEG). One requirement for this understanding is a \emph{forward model} that links the `hidden' activities of billions of neurons in mammalian brains and their propagation through neural networks to experimentally accessible quantities such as LFP and EEG. Utilizing terminology from theoretical physics, we call the operationally accessible quantities \emph{observables} and an integrative forward model an \emph{observation model}. Moreover, there is an ongoing debate in the literature whether field effects, i.e. the feedback from mass potentials to neural activity, plays a functional role in the self-organization of cortical activity (e.g. \cite{Jefferys95}). Such field effects have recently been demonstrated via experiments on ephatic interaction \cite{FrohlichMcCormick10}. Thus a theoretical framework for observation models is mandatory in order to describe that coupling in clinical, computational and cognitive neurosciences, e.g. for treatment of epilepsy \cite{RichardsonSchiffGluckman05} or modeling cognition-related brain potentials \cite{GrabenPotthast12a, LaszloPlaut12}.

\begin{figure}[H]
 \centering
\includegraphics[width=0.95\textwidth]{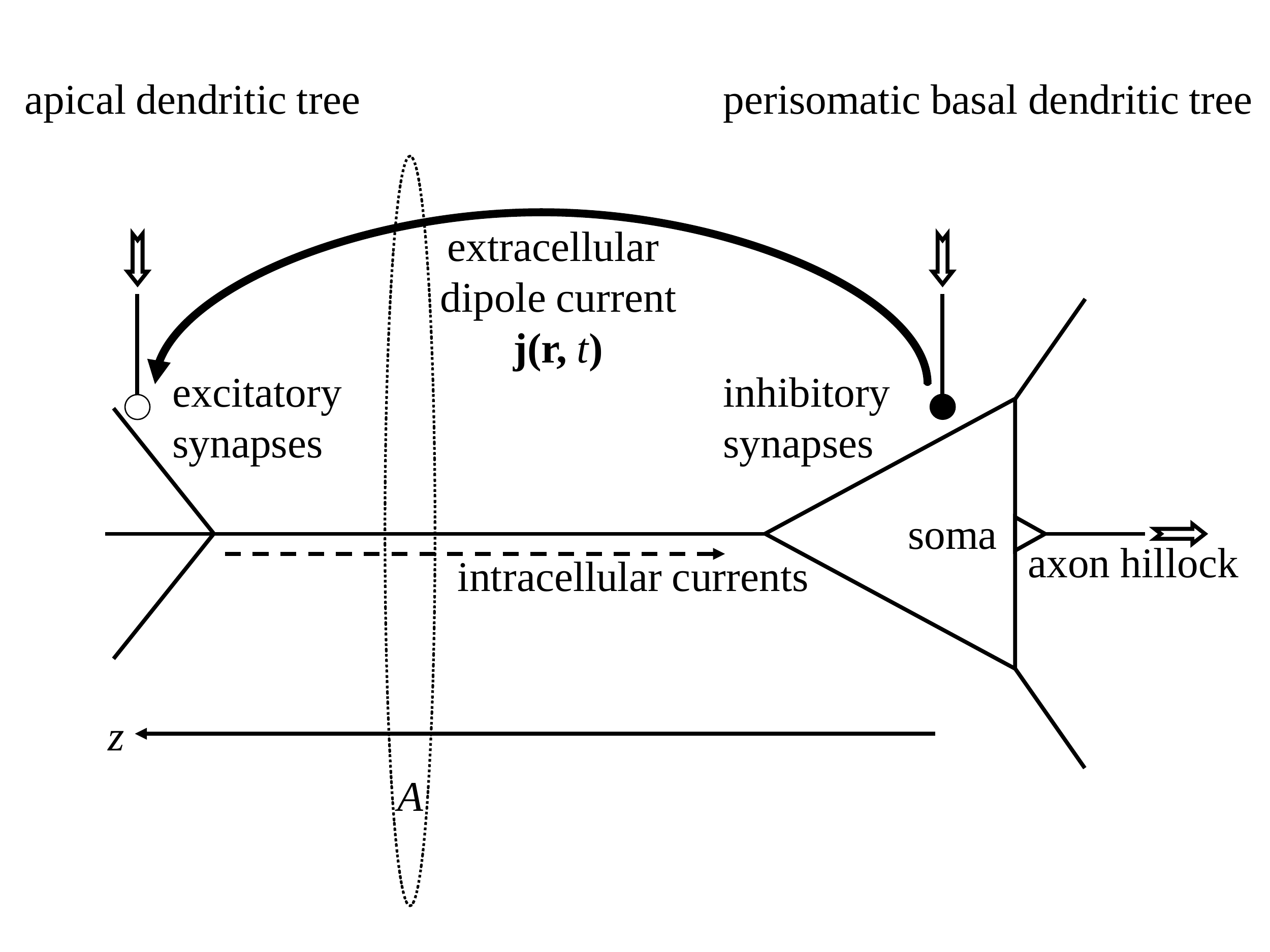}
\caption{\label{fig:GS:pyramid} Sketch of a cortical pyramidal neuron with extracellular current dipole between spatially separated excitatory (open bullet) and inhibitory synapses (filled bullet). Neural in- and outputs are indicated by the jagged arrows. The $z$-axis points toward the scull. Current density $j$ is given by dendritic current $I_1$ through cross section area $A$ as described in the text.}
\end{figure}

Currently, there is ample evidence that the generators of neural field potentials, such as cortical LFP and EEG are the cortical pyramidal cells (sketched in \Fig{fig:GS:pyramid}). They exhibit a long dendritic trunk separating mainly excitatory synapses at the apical dendritic tree from mainly inhibitory synapses at the perisomatic basal dendritic tree \cite{CreutzfeldtWatanabeLux66a, Spruston08}. When both kinds of synapses are simultaneously active, inhibitory synapses generate current sources and excitatory synapses current sinks in extracellular space, causing the pyramidal cell to behave as a microscopic dipole surrounded by its characteristic electrical field. This dendritic dipole field is conveniently described by its associated electrodynamic potential, the \emph{dendritic field potential (DFP)}. Dendritic fields superimpose to the field of a cortical dipole layer, which is measurable as cortical LFP, due to the geometric arrangement of pyramidal cells in a cortical column.  There pyramidal cells exhibit an axial symmetry and are aligned in parallel to each other, perpendicular to the cortex' surface, thus forming a palisade of cell bodies and dendritic trunks. Eventually, cortical LFP gives rise to the EEG measurable at the human's scalp \cite{DestexheBedard13, NunezSrinivasan06, SchomerSilva11}.

Weaving the above phenomena into a mathematical and biophysical plausible observation model that represents correctly the multi-spatiotemporal characteristics of LFP is a non-trivial task. The difficulty results from the complexity of brain processes that operate at several spatial and temporal scales. On one hand the organization of the brain, from single neuron scales to that of whole brain regions, changes its connectivity from almost probabilistic to highly structured as discussed above in the case of the cortical columns. On the other hand, temporal dynamics in time scales ranges from milliseconds for discrete events like spikes to hours and even longer for synaptic plasticity and learning. Hence, there is strong interaction between the different spatiotemporal scales \cite{Berry1997, lakatos2005oscillatory}, which directly contribute to complex oscillatory dynamics, e.g., to \emph{mixed-mode oscillations} \cite{IrinaDavid2008, DesrochesEtAl2012}. Thus it is not clear how and when to break down complex brain processes into simpler `building blocks' where analysis can be made. Despite these peculiarities, various mathematical and computational approaches have been proposed in order to establish coarse-graining techniques and how to move from one scale to another.

Most studies for realistically simulating LFP, typically for the extracellular space in the vicinity of a neuron, have been attempted by means of compartmental models \cite{BazhenovEA11, LindenTetzlaffEA11, PettersenHagenEinevoll08, SargsyanPapatheodoropoulosKostopoulos01} where every compartment contributes a portion of extracellular current to the DFP that is given by Coulomb's equation in conductive media \cite{BedardKroegerDestexhe04, Graben08a, NunezSrinivasan06}. However, because compartmental models are computationally expensive, large-scale neural networks preferentially employ point models, based either on conductance \cite{HodgkinHuxley52, MazzoniPanzeriEA08} or population models \cite{JansenRit95, rodrigues2010mappings, WendlingBellangerEA00, WilsonCowan72} where neural mass potentials are estimated either through sums (or rather differences) of postsynaptic potentials \cite{DavidFriston03} or of postsynaptic currents \cite{MazzoniPanzeriEA08}. In particular, the model of Mazzoni et al. \cite{MazzoniPanzeriEA08} led to a series of recent follow-up studies \cite{MazzoniEA10, MazzoniBrunelEA11} that address the correlations between numerically simulated or experimentally measured LFP/EEG and spike rates by means of statistical modeling and information theoretic measures.

To adequately explain field potentials measured around the dendritic tree of an individual cortical pyramidal cell (DFP), in extracellular space of a cortical module (LFP), or at a human's scalp (EEG), Maxwell's electromagnetic field equations, specifically the continuity equation describing conservation of charge has to be taken into account. However, coupling the activity of discrete neural networks to the continuous electromagnetic field is difficult since neural network topology is not embedded into physical space as an underlying metric manifold. This can be circumvented by employing continuous neural networks as investigated in \emph{neural field theory (NFT)} \cite{Amari77b, BreakspearRobertsEA06, Bressloff12, HuttAtay05, JirsaHaken97, WilsonCowan73}. In fact previous studies \cite{JirsaEA02, LileyCaduschDafilis02} gave the first reasonable accounts for such couplings in NFT population models that are motivated by the corresponding assumptions for neural mass models (cf. the chapter of Pinotsis and Friston in this volume). Jirsa et al. \cite{JirsaEA02} relate the impressed current density in extracellular space to neural field activity. On the other hand, Liley et al. \cite{LileyCaduschDafilis02} consider LFP as average somatic membrane potential being proportional to the neural field. Their model found a number of successful applications \cite{BojakLiley05, CoombesEA07, BojakOostendorpEA10} (see also the chapter of Liley in this volume). However, both approaches \cite{JirsaEA02, LileyCaduschDafilis02} are not concerned with the microscopic geometry around the field generators, the cortical pyramidal cells. Therefore, they do not take pyramidal dipole currents into account.

Another problem with the aforementioned neural field approaches is that the extracellular space has either been completely neglected, or only implicitly been taken into account by assuming that cortical LFP is proportional to either membrane potentials or synaptic currents as resulting from a purely resistive medium. That means, dipole currents in the extracellular space have been completely abandoned. However, recent studies indicate that at least the resistive property of the extracellular space is crucial \cite{Logothetis2007809}, but more interestingly, it has been revealed that diffusion currents, represented by their corresponding Warburg impedances \cite{SkaleDolecekSlemnik07}, cannot be neglected in extracellular space as they may substantially contribute to the characteristic power spectra of neural mass potentials \cite{BedardDestexhe09, BedardRodriguesEA10, BedardDestexhe12, DestexheBedard13}.

In this chapter, we outline a theoretical framework for the microscopic coupling of continuous neural networks, i.e. neural fields, to the electromagnetic field, properly described by dipole currents of cortical pyramidal neurons and diffusion effects in extracellular space. As a starting point we use a three-compartment model for a single pyramidal cell \cite{Destexhe01, Graben08a, WangTegnerEA04} and derive the evolution law for the activity of a neural network. These derivations additionally include observation equations for the extracellular dipole currents, which explicitly incorporate extracellular resistivity and diffusion. Subsequently, we demonstrate that our approach can be related to previous modeling strategies, by considering reasonable simplifications. Herein, we intentionally and specifically simplify our approach to a leaky integrate-and-fire (LIF) model for the dynamics of a neural network, which then shows the missing links that previous modeling approaches failed to incorporate to account for a proper dipole LFP observation model. In particular, we compare our results with the related model by Mazzoni et al. \cite{MazzoniPanzeriEA08} by means of numerical simulations. Moreover, performing the continuum limit (yet \emph{\`{a} la physique}) for the network yields an Amari-type neural field equation \cite{Amari77b} coupled to the Maxwell equations in extracellular fluid, while an observation model for electric field potentials is obtained from the interaction of cortical dipole currents with charge density in non-resistive extracellular space as described by the Nernst-Planck equation. Thereby, our work provides for the first time a biophysically plausible observation model for the Amari-type neural field equations and crucially, it gives estimates for the local field potentials that satisfy the widespread dipole assumption discussed in the neuroscientific literature.


\section{Pyramidal Neuron Model}
\label{sec:GS:3comp}

\begin{landscape}
\begin{figure}[H]
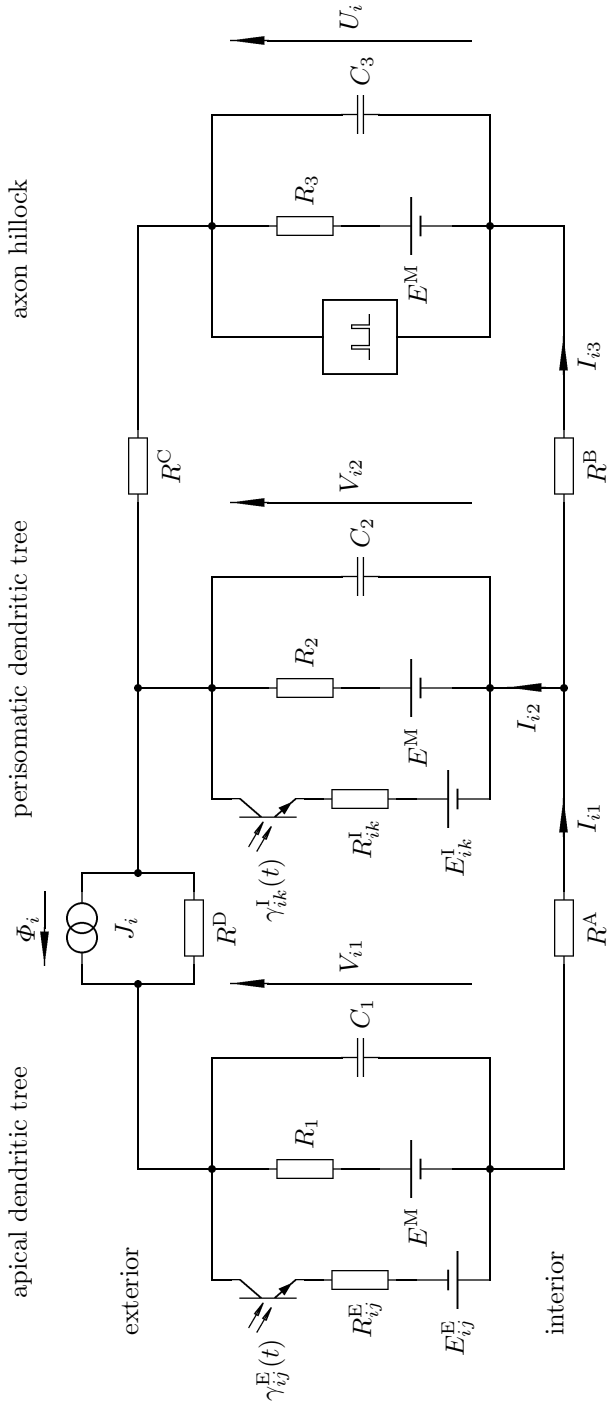

\centering
\begin{circuit}0
\.1
\- 4 d
\.2
\- 3 d
\R2 {} d
\- 2 d
\nl \U1 {$E^\mathrm{M}$} + d
\- 2 d
\.3
\- 4 d
\.4
\atpin .2
\- 6 l
\- 1 d
\nl \npnPH1 {$\gamma_{ik}^\mathrm{I}(t)$} C D
\frompin npnPH1E
\- 1 d
\nl \R5 {$R_{ik}^\mathrm{I}$} d
\- 1 d
\nl \U2 {$E_{ik}^\mathrm{I}$} + d
\- 6 r
\atpin .2
\- 6 r
\- 7 d
\vcenterto R5
\C2 {} d
\- 6 d
\- 6 l
\atpin .2: \shift 10 0 \P1
\atpin .3: \shift 10 0 \P2
\Utext {$V_{i2}$} from P2 to P1
%
%
\atpin .1
\- 10 l
\.5
\- 3 u
\- 1 l
\nl \I1 {$J_i$} l
\- 1 l
\- 3 d
\.6
\- 3 d
\- 1 r
\nl \R6 {$R^\mathrm{D}$} r
\- 1 r
\- 3 u
\atpin .5: \shift 0 5 \P7
\atpin .6: \shift 0 5 \P8
\Utext {$\Phi_i$} from P7 to P8
%
%
\atpin .6
\- 10 l
\- 4 d
\.7
\- 3 d
\R1 {} d
\- 2 d
\nl \U3 {$E^\mathrm{M}$} + d
\- 2 d
\.8
\- 4 d
\atpin .7
\- 6 l
\- 1 d
\nl \npnPH2 {$\gamma_{ij}^\mathrm{E}(t)$} C D
\frompin npnPH2E
\- 1 d
\nl \R4 {$R_{ij}^\mathrm{E}$} d
\- 1 d
\nl \U4 {$E_{ij}^\mathrm{E}$} - d
\- 6 r
\atpin .7
\- 6 r
\- 7 d
\vcenterto R4
\C1 {} d
\- 6 d
\- 6 l
\atpin .7: \shift 10 0 \P3
\atpin .8: \shift 10 0 \P4
\Utext {$V_{i1}$} from P4 to P3
%
%
\atpin .8
\- 4 d
\- 11 r
\hcenterto R6
\nl \R7 {$R^\mathrm{A}$} r
\- 5 r
\nl \whatI1 {$I_{i1}$} d l
\- 7 r
%
%
\- 11 r
\nl \R8 {$R^\mathrm{B}$} r
\- 5 r
\nl \whatI2 {$I_{i3}$} d l
\- 8 r
\atpin .4
\moverel{0} {1}
\nl \whatI3 {$I_{i2}$} s u
%
%
\atpin .1
\- 10 r
\hcenterto R8
\nl \R10 {$R^\mathrm{C}$} r
\- 11 r
%
%
\- 4 d
\.9
\- 3 d
\R3 {} d
\- 2 d
\nl \U5 {$E^\mathrm{M}$} + d
\- 2 d
\.10
\- 4 d
\atpin .9
\- 6 l
\- 1 d
\vcenterto R4
\Impulse1 P4 {} {} {} {}
\frompin Impulse1P4
\- 5 u
\frompin Impulse1P2
\- 5 d
\- 6 r
\atpin .9
\- 6 r
\- 7 d
\vcenterto R4
\C3 {} d
\- 6 d
\- 6 l
\atpin .9: \shift 10 0 \P5
\atpin .10: \shift 10 0 \P6
\Utext {$U_i$} from P6 to P5
%
%
\atpin .7: \shift -7 10
\put{apical dendritic tree}
\atpin .2: \shift -7 10
\put{perisomatic dendritic tree}
\atpin .9: \shift -5 10
\put{axon hillock}
\atpin .1: \shift -35 0
\put{exterior}
\atpin .4: \shift -35 0
\put{interior}
\end{circuit}
\caption{\label{fig:GS:neuroncirc} Equivalent circuit for a three-compartment pyramidal neuron model.}
\end{figure}
\end{landscape}

Inspired by earlier attempts to describe the complex shape of cortical pyramidal neurons by essentially three passively coupled compartments \cite{Destexhe01, Graben08a, WangTegnerEA04}, we reproduce in \Fig{fig:GS:neuroncirc} the electronic equivalent circuit of beim Graben \cite{Graben08a} for the $i$th pyramidal cell [\Fig{fig:GS:pyramid}] in a population of $P$ pyramidal neurons here. This parsimonious circuit allows to derive our observation model. It comprises one compartment for the apical dendritic tree where $p_i$ excitatory synapses are situated (for the sake of simplicity, we only show one synapse here), another one for the soma and peri\-somatic basal dendritic tree, populated with $q_i$ mainly inhibitory synapses (again, only one synapse is shown here), and a third one for the axon hillock where membrane potential is converted into spike trains by means of an integrate-and-fire mechanism. Note that nonlinear fire mechanisms of Hodgkin-Huxely type can be incorporated as well. In total we consider $N$ populations of neurons, arranged in two-dimensional layers $\Gamma_n \subset \mathbb{R}^2$ ($i = n, \dots, N$). Neurons in layers $1$ to $M$ should be excitatory, neurons in layers $M+1$ to $N$ should be inhibitory and layer 1 exclusively contains the $P$ cortical pyramidal cells in our simplified treatment. The total number of neurons should be $K$.

Excitatory synapses are schematically represented by the left-most branch of \Fig{fig:GS:neuroncirc} as `phototransistors' \cite{Graben08a} in order to indicate that they comprise quanta-gated resistors, namely ion channels whose resistance depends on the concentration of ligand molecules which are either extracellular neurotransmitters or intracellular metabolites \cite{KandelSchwartzEA91}. There, the excitatory postsynaptic current (EPSC) at a synapse between a neuron $j$ from layers $1$ to $M$ and neuron $i$ is given as
\begin{equation}\label{eq:GS:epsc}
    I_{ij}^\mathrm{E}(t) = \frac{\gamma_{ij}^\mathrm{E}(t)}{R_{ij}^\mathrm{E}}(V_{i1}(t) - E_{ij}^\mathrm{E}) \:.
\end{equation}
Here, the time-dependent function $\gamma_{ij}^\mathrm{E}(t)$ reflects the neurotransmitter-gated opening of postsynaptic ion channels. Usually, this function is given as a sum of characteristic excitatory impulse response functions $\eta^\mathrm{E}(t)$ that is elicited by one presynaptic spike, i.e.
\begin{equation}\label{eq:GS:gamma}
   \gamma(t) = \sum_\nu \eta(t - t_\nu)
\end{equation}
where $t_\nu$ denote the ordered spike arrival times. Moreover, $R_{ij}^\mathrm{E}$ comprises the maximum synaptic conductivity as well as the electrotonic distance between the synapse between neuron $j$ and $i$ and $i$'s trigger zone, both expressed as resistance. $V_{i1}(t)$ is the membrane potential of neuron $i$'s compartment 1, i.e. the apical dendritic tree and $E_{ij}^\mathrm{E}$ is the excitatory reversal potential of the synapse $j \to i$. We can conveniently express $\gamma(t)$ through the spike rate \cite{Bressloff12, GrabenLiebscherKurths08}
\begin{equation}\label{eq:GS:srate}
   a(t) = \sum_\nu \delta(t - t_\nu)
\end{equation}
by means of a temporal convolution (`$*$' denotes convolution product)
\begin{equation}\label{eq:GS:conv}
   \gamma(t) = \int_{-\infty}^t \eta(t - t') a(t') \, \D t' = (\eta * a)(t) \:.
\end{equation}
Furthermore, the apical dendritic compartment 1 is characterized by a particular leakage resistance $R_1$ and by a capacity $C_1$, reflecting the temporary charge storage capacity of the membrane. Both, $R_1$ and  $C_1$ are correlated with the compartment's membrane area \cite{Destexhe01}. The battery $E^\mathrm{M}$ denotes the Nernst resting potential \cite{Tuckwell88a, JohnstonWu97}.

The middle branch of \Fig{fig:GS:neuroncirc} describes the inhibitory synapses (also displayed as `phototransistors' \cite{Graben08a}) between a neuron $k$ from layers $M+1$ to $N$ and neuron $i$. Here, inhibitory postsynaptic currents (IPSC)
\begin{equation}\label{eq:GS:ipsc}
    I^\mathrm{I}_{ik}(t) = \frac{\gamma_{ik}^\mathrm{I}(t)}{R_{ik}^\mathrm{I}}(V_{i2}(t) - E_{ik}^\mathrm{I}) \:,
\end{equation}
described by a similar channel opening function $\gamma^\mathrm{I}(t)$, shunting the excitatory branch with the trigger zone when compartment's 2 membrane potential $V_{i2}(t)$ is large due to previous excitation. Also Eqs. \pref{eq:GS:gamma} and \pref{eq:GS:srate} hold for another postsynaptic impulse response function $\eta^\mathrm{I}(t)$, characteristic for inhibitory synapses. The resistance of the current paths along the cell plasma is given by $R^\mathrm{I}_{ik}$. Finally,  $E_{ik}^\mathrm{I}$ denotes the inhibitory reversal potential of the synapse $k \to i$. Also the somatic and perisomatic dendritic compartment 2 possesses its specific leakage resistance $R_2$ and capacity $C_2$.

The cell membrane at the axon hillock \cite{HodgkinHuxley52} itself is represented by the branch at the right hand side described by another RC-element consisting of $R_3$ and $C_3$. Action potentials, $\delta{(t-t_{\nu})}$, are generated by a leaky integrate-and-fire mechanism \cite{MazzoniPanzeriEA08} as indicated by a `black box' when the membrane potential $U_i(t)$ crosses a certain threshold $\theta_i > 0$ from below at time $t_\nu$, i.e.
\begin{equation}\label{eq:GS:thres}
     U_i(t_\nu) \ge \theta_i \:.
\end{equation}
Afterwards, membrane potential is reset to some steady-state potential \cite{MazzoniPanzeriEA08}
\begin{equation}\label{eq:GS:reset}
 U_i(t_{\nu+1}) \leftarrow E \:.
\end{equation}
and the integration of the differential equations can be restarted at time $t = t_{\nu +1} + \tau_{rp}$ after interrupting the dynamics for a refractory period $\tau_{rp}$.

The three compartments are coupled through longitudinal resistors, $R^\mathrm{A}$, $R^\mathrm{B}$, $R^\mathrm{C}$, $R^\mathrm{D}$ where $R^\mathrm{A}, R^\mathrm{B}$ denote the resistivity of the cell plasma and $R^\mathrm{C}, R^\mathrm{D}$ that of extracellular space \cite{HoltKoch99a}. Yet, in extracellular space not only Ohmic but also diffusion currents are present \cite{BedardDestexhe09, BedardRodriguesEA10, BedardDestexhe12, GabrielEA96a, GabrielEA96b , GabrielEA96c, SykovaNicholson08}. These are taken into account by the current source $J_i$ connected in parallel to $R^\mathrm{D}$. However, for convenience, diffusion currents in the extracellular space  between compartments 2 and 3 are disregarded following an adiabatic approximation but somatic resistor $R^\mathrm{C}$ is accounted for.

Finally, the membrane potentials at compartments 1, $V_{i1}$, 2, $V_{i2}$, and 3, $U_i$ as the dynamical state variable as well as the DFP $\Phi_i$ are shown in \Fig{fig:GS:neuroncirc}. The latter drops along the extracellular resistor $R^\mathrm{D}$. For the aim of calculation, the mesh currents $I_{i1}$ (current in the apical compartment 1 of neuron $i$), $I_{i2}$ (current in somatic and perisomatic compartment 2 of neuron $i$) and $I_{i3}$ (the leaky integrate-and-fire (LIF) current in compartment 3 of neuron $i$) are indicated.

The circuit in \Fig{fig:GS:neuroncirc} obeys Kirchhoff's laws; first for currents:
\begin{eqnarray}
 \label{eq:GS:circuit1}
    I_{i1} &=& C_1 \frac{\D V_{i1}}{\D t} + \frac{V_{i1} - E^\mathrm{M}}{R_1} + \sum_{j = 1}^{p_i} I_{ij}^\mathrm{E} \\
  \label{eq:GS:circuit2}
    I_{i2} &=& C_2 \frac{\D V_{i2}}{\D t} + \frac{V_{i2} - E^\mathrm{M}}{R_2} + \sum_{k = 1}^{q_i} I_{ik}^\mathrm{I} \\
  \label{eq:GS:circuit3}
    I_{i3} &=& C_3 \frac{\D U_i}{\D t} + \frac{U_i - E^\mathrm{M}}{R_3} \\
  \label{eq:GS:circuit4}
    I_{i3} &=& I_{i1} -  I_{i2} \:,
\end{eqnarray}
and second, for voltages:
\begin{eqnarray}
 \label{eq:GS:circuit5}
    V_{i1} &=&  (R^\mathrm{A} + R^\mathrm{D}) I_{i1}  +  (R^\mathrm{B} + R^\mathrm{C}) I_{i3} + U_i -  R^\mathrm{D} J_i \\
 \label{eq:GS:circuit6}
    V_{i2} &=&  (R^\mathrm{B} + R^\mathrm{C}) I_{i3} + U_i \\
 \label{eq:GS:circuit7}
   \Phi_i &=& R^\mathrm{D} (I_{i1} - J_i) \:,
\end{eqnarray}
where the postsynaptic currents $I_{ij}^\mathrm{E}$ and $I_{ik}^\mathrm{I}$ are given through \pref{eq:GS:epsc} and \pref{eq:GS:ipsc}. Here, $p_i$ is the number of excitatory and $q_i$ the number of inhibitory synapses connected to neuron $i$.

Subtracting \pref{eq:GS:circuit6} from \pref{eq:GS:circuit5} yields the current along the pyramidal cell's dendritic trunk
\begin{equation}\label{eq:GS:dendcurrent}
    I_{i1} = \frac{V_{i1} - V_{i2} + R^\mathrm{D} J_i}{R^\mathrm{A} + R^\mathrm{D}} \:.
\end{equation}
The circuit described by Eqs. (\ref{eq:GS:circuit1} -- \ref{eq:GS:circuit7}) shows that the neuron $i$ is likely to fire when the excitatory synapses are activated. Then, the LIF current $I_{i3}$ equals the dendritic current $I_{i1}$. If, by contrast, also the inhibitory synapses are active, the dendritic current $I_{i1}$ follows the shortcut between the apical and perisomatic dendritic trees and only a portion could evoke spikes at the trigger zone [\Eq{eq:GS:circuit3}]. On the other hand, the large dendritic current $I_{i1}$, diminished by some diffusion current $J_i$, flowing through the extracellular space of resistance $R^\mathrm{D}_i$, gives rise to a large DFP $\Phi_i$.

In order to simplify the following derivations, we first gauge the resting potential to $E^\mathrm{M} = 0$. Then, excitatory synapses are characterized by $E_{ij}^\mathrm{E} > 0$, while inhibitory synapses obey $E_{ik}^\mathrm{I} < 0$. Combining Eqs. (\ref{eq:GS:circuit1} -- \ref{eq:GS:circuit6}) entails
\begin{eqnarray}
  \label{eq:GS:circuit1a}
    C_1 \frac{\D V_{i1}}{\D t} + \frac{V_{i1}}{R_1} + \sum_{j = 1}^{p_i} I_{ij}^\mathrm{E} &=&
        \frac{V_{i1} - V_{i2} + R^\mathrm{D} J_i}{R^\mathrm{A} + R^\mathrm{D}}    \\
  \label{eq:GS:circuit2b}
    C_2 \frac{\D V_{i2}}{\D t} + \frac{V_{i2}}{R_2} + \sum_{k = 1}^{q_i} I_{ik}^\mathrm{I} &=&
    \frac{V_{i1} - V_{i2} + R^\mathrm{D} J_i}{R^\mathrm{A} + R^\mathrm{D}} - \frac{V_{i2} - U_i}{R^\mathrm{B} + R^\mathrm{C}} \\
  \label{eq:GS:circuit3c}
    C_3 \frac{\D U_i}{\D t} + \frac{U_i}{R_3} &=&
        \frac{V_{i2} - U_i}{R^\mathrm{B} + R^\mathrm{C}} \:.
\end{eqnarray}


\subsection{General Solution}
\label{sec:GS:gensol}

Next we follow Bressloff's \cite{Bressloff94} argumentation and regard the compartmental voltages as auxiliary variables that are merged into a two-dimensional vector $\vec{V}_i = (V_{i1}, V_{i2})^T$ which is subject to elimination. We only keep \Eq{eq:GS:circuit3c} as the evolution law of the entire state vector $\vec{U} = (U_i)_{i=1,\dots, K}$ of the neural network. Inserting the postsynaptic currents from \pref{eq:GS:epsc} and \pref{eq:GS:ipsc} into Eqs. (\ref{eq:GS:circuit1a}, \ref{eq:GS:circuit2b}) and temporarily assuming an arbitrary time-dependence for the functions $\gamma(t)$ from \Eq{eq:GS:gamma} (in fact, the $\gamma(t)$ are given through the presynaptic spike rates and are thus nonlinear functions of the entire state $\vec{U}$), we obtain a system of two inhomogeneous linear differential equations that can be compactly written in matrix form as
\begin{equation}\label{eq:GS:cmpvode}
    \frac{\D}{\D t} \vec{V}_i(t) = \vec{H}_i(t) \cdot \vec{V}_i(t) + \vec{G}_i(t) \:,
\end{equation}
with
\begin{equation}\label{eq:GS:hmat}
    \vec{H}_i(t) =
    \begin{pmatrix}
      \frac{1}{C_1} \left(
      -\frac{1}{R_1} + \frac{1}{R^\mathrm{A} + R^\mathrm{D}} - \sum_j \frac{\gamma_{ij}^\mathrm{E}(t)}{R_{ij}^\mathrm{E}}
      \right)
      &
      -\frac{1}{C_1 (R^\mathrm{A} + R^\mathrm{D})} \\
      \frac{1}{C_2 (R^\mathrm{A} + R^\mathrm{D})}
      &
      \frac{1}{C_2}\left(
      -\frac{1}{R_2} - \frac{1}{R^\mathrm{A} + R^\mathrm{D}} - \frac{1}{R^\mathrm{B} + R^\mathrm{C}}
      - \sum_k \frac{\gamma_{ik}^\mathrm{I}(t)}{R_{ik}^\mathrm{I}} \right)
    \end{pmatrix}
\end{equation}
and
\begin{equation}\label{eq:GS:gmat}
    \vec{G}_i(t) =
    \begin{pmatrix}
        \sum_j \frac{\gamma_{ij}^\mathrm{E}(t) E_{ij}^\mathrm{E}}{C_1 R_{ij}^\mathrm{E}}  +
        \frac{R^\mathrm{D}}{C_1 (R^\mathrm{A} + R^\mathrm{D})} J_i(t) \\
        \sum_k \frac{\gamma_{ik}^\mathrm{I}(t) E_{ik}^\mathrm{I}}{C_2 R_{ik}^\mathrm{I}}  +
        \frac{R^\mathrm{D}}{C_2 (R^\mathrm{A} + R^\mathrm{D})} J_i(t) + \frac{1}{C_2 (R^\mathrm{B} + R^\mathrm{C})} U_i(t)
    \end{pmatrix} \:.
\end{equation}
As initial conditions start with $\vec{V}  = 0$ in the infinite past $t = -\infty$ for the sake of convenience.

Obviously, the time-dependence of the transition matrix $\vec{H}(t)$ is due to the shunting terms $\gamma(t)$. In order to solve \pref{eq:GS:cmpvode} one first considers the associated homogeneous differential equation
\begin{equation}\label{eq:GS:cmpvodeh}
    \frac{\D}{\D t} \vec{W}_i(t) = \vec{H}_i(t) \cdot \vec{W}_i(t)
\end{equation}
whose general solutions are given as the columns of
\begin{equation}\label{eq:GS:solvhomde}
    \vec{W}_i(t) = \TO \E^{t \vec{H}_i(t)} \:,
\end{equation}
where $\TO$ denotes the \emph{time-ordering operator} \cite{BressloffTaylor93b, BressloffTaylor93a}. Using the integral \pref{eq:GS:solvhomde}, a particular solution of the inhomogeneous equation \pref{eq:GS:cmpvode} is then obtained by the variation of parameter method as
\begin{equation}\label{eq:GS:solvinhode}
    \vec{V}_i(t) = \int_{-\infty}^t \vec{X}_i(t, t') \cdot \vec{G}_i(t') \, \D t'
\end{equation}
with matrix-valued Green's function
\begin{equation}\label{eq:GS:green}
    \vec{X}_i(t, t') = \vec{W}_i(t) \cdot \vec{W}_i(t')^{-1} \:.
\end{equation}
Therefore, the compartmental voltages are obtained as
\begin{equation}\label{eq:GS:solvinhodecom}
    V_{i\alpha}(t) = \sum_{\beta = 1}^2 \int_{-\infty}^t \chi_{i \alpha \beta}(t, t') g_{i\beta}(t') \, \D t' =
     \sum_{\beta = 1}^2 \chi_{i \alpha \beta} * g_{i\beta}
\end{equation}
with components $\vec{X}_i(t, t') = (\chi_{i \alpha \beta}(t, t'))_{\alpha \beta}$ and $\vec{G}_i(t') = (g_{i\beta}(t'))_\beta$, $\alpha, \beta = 1, 2$.


\subsection{Observation Model}
\label{sec:GS:obsmod}

In order to derive the general observation equations for the DFP of the three-compartment model, we insert the formal solutions \pref{eq:GS:solvinhodecom} and the inhomogeneity \pref{eq:GS:gmat} into \Eq{eq:GS:dendcurrent} and obtain
\begin{multline}\label{eq:GS:dendcurrent1}
        I_{i1}(t) = \frac{1}{R^\mathrm{A} + R^\mathrm{D}} \int_{-\infty}^t
    (\chi_{i 11}(t, t') - \chi_{i 21}(t, t'))
    \Bigg[
    \sum_j \frac{E_{ij}^\mathrm{E}}{C_1 R_{ij}^\mathrm{E}} \gamma_{ij}^\mathrm{E}(t)  +
        \frac{R^\mathrm{D}}{C_1 (R^\mathrm{A} + R^\mathrm{D})} J_i(t)
        \Bigg] + \\
    (\chi_{i 12}(t, t') - \chi_{i 22}(t, t'))
    \Bigg[
      \sum_k \frac{E_{ik}^\mathrm{I}}{C_2 R_{ik}^\mathrm{I}} \gamma_{ik}^\mathrm{I}(t) +
        \frac{R^\mathrm{D}}{C_2 (R^\mathrm{A} + R^\mathrm{D})} J_i(t) + \frac{1}{C_2 (R^\mathrm{B} + R^\mathrm{C})} U_i(t)
        \Bigg] \, \D t'+ \\
        \frac{R^\mathrm{D}}{R^\mathrm{A} + R^\mathrm{D}} J_i(t)  \:,
\end{multline}
which can be reshaped by virtue of the convolutions \pref{eq:GS:conv} to
\begin{multline}\label{eq:GS:dendcurrent2}
    I_{i1} = \frac{1}{R^\mathrm{A} + R^\mathrm{D}}
    \Bigg[
    \sum_j \frac{E_{ij}^\mathrm{E}}{C_1 R_{ij}^\mathrm{E}} (\chi_{i 11} - \chi_{i 21}) * \eta^\mathrm{E} * a_j  +
        \frac{R^\mathrm{D}}{C_1 (R^\mathrm{A} + R^\mathrm{D})} (\chi_{i 11} - \chi_{i 21}) * J_i
        \Bigg] + \\
    \frac{1}{R^\mathrm{A} + R^\mathrm{D}}
    \Bigg[
      \sum_k \frac{E_{ik}^\mathrm{I}}{C_2 R_{ik}^\mathrm{I}} (\chi_{i 12} - \chi_{i 22}) * \eta^\mathrm{I} * a_k +
        \frac{R^\mathrm{D}}{C_2 (R^\mathrm{A} + R^\mathrm{D})} (\chi_{i 12} - \chi_{i 22}) * J_i  + \\
        \frac{1}{C_2 (R^\mathrm{B} + R^\mathrm{C})} (\chi_{i 12} - \chi_{i 22}) * U_i \Bigg] +
        \frac{R^\mathrm{D}}{R^\mathrm{A} + R^\mathrm{D}} J_i  \:.
\end{multline}

Introducing new impulse response functions that simultaneously account for synaptic transmission ($\eta$) and dendritic propagation ($\chi$) by
\begin{eqnarray}
    \label{eq:GS:green2a}
        \psi_{i \alpha 1} &=&  \chi_{i \alpha 1} * \eta^\mathrm{E} \\
    \label{eq:GS:green2b}
        \psi_{i \alpha 2} &=&  \chi_{i \alpha 2} * \eta^\mathrm{I}
\end{eqnarray}
yields
\begin{multline}\label{eq:GS:dendcurrent3}
    I_{i1} = \frac{1}{R^\mathrm{A} + R^\mathrm{D}}
    \Bigg[
    \sum_j \frac{E_{ij}^\mathrm{E}}{C_1 R_{ij}^\mathrm{E}} (\psi_{i 11} - \psi_{i 21}) * a_j  +
        \frac{R^\mathrm{D}}{C_1 (R^\mathrm{A} + R^\mathrm{D})} (\chi_{i 11} - \chi_{i 21}) * J_i
        \Bigg] + \\
    \frac{1}{R^\mathrm{A} + R^\mathrm{D}}
    \Bigg[
      \sum_k \frac{E_{ik}^\mathrm{I}}{C_2 R_{ik}^\mathrm{I}} (\psi_{i 12} - \psi_{i 22}) * a_k +
        \frac{R^\mathrm{D}}{C_2 (R^\mathrm{A} + R^\mathrm{D})} (\chi_{i 12} - \chi_{i 22}) * J_i  + \\
        \frac{1}{C_2 (R^\mathrm{B} + R^\mathrm{C})} (\chi_{i 12} - \chi_{i 22}) * U_i \Bigg]  + \frac{R^\mathrm{D}}{R^\mathrm{A} + R^\mathrm{D}} J_i \:.
\end{multline}
Eventually we obtain the DFP of neuron $i$ as the potential dropping along the resistor $R^\mathrm{D}$ caused by the current through it [\Eq{eq:GS:circuit7}], i.e.
\begin{multline}\label{eq:GS:dfp}
  \Phi_i =
  \frac{R^\mathrm{D}}{R^\mathrm{A} + R^\mathrm{D}}
    \Bigg\{
    \sum_j \frac{E_{ij}^\mathrm{E}}{C_1 R_{ij}^\mathrm{E}} (\psi_{i 11} - \psi_{i 21}) * a_j  +
    \sum_k \frac{E_{ik}^\mathrm{I}}{C_2 R_{ik}^\mathrm{I}} (\psi_{i 12} - \psi_{i 22}) * a_k + \\
        \Bigg[ \frac{R^\mathrm{D}}{C_1 (R^\mathrm{A} + R^\mathrm{D})} (\chi_{i 11} - \chi_{i 21}) +
                  \frac{R^\mathrm{D}}{C_2 (R^\mathrm{A} + R^\mathrm{D})} (\chi_{i 12} - \chi_{i 22}) - R^\mathrm{A} \delta
        \Bigg]* J_i  + \\
        \frac{1}{C_2 (R^\mathrm{B} + R^\mathrm{C})} (\chi_{i 12} - \chi_{i 22}) * U_i  \Bigg\}
\end{multline}


\subsection{Neurodynamics}
\label{sec:GS:neudy}

Equation \pref{eq:GS:dfp} reveals that the DFP is driven by the neuron's state variable $U_i$, by the entirety of postsynaptic potentials caused by spike trains $a_i$ and by the diffusion currents $J_i$. The state variables and the spike trains are given by the network's evolution equation that is straightforwardly derived along the lines of Bressloff \cite{Bressloff94} again. To this end, we insert $V_{i 2}(t)$ as the solution of \pref{eq:GS:solvinhodecom} into the remaining equation \pref{eq:GS:circuit3c} to get
\begin{multline}\label{eq:GS:nd}
    C_3 (R^\mathrm{B} + R^\mathrm{C}) \frac{\D U_i}{\D t} + \left( 1 + \frac{R^\mathrm{B} + R^\mathrm{C}}{R_3} \right) U_i =
     \chi_{i 21} *  g_{i1} + \chi_{i 22} *  g_{i2} \:.
\end{multline}
Next, we insert the inhomogeneity \pref{eq:GS:gmat} again and obtain
\begin{multline}\label{eq:GS:nd1}
    C_3 (R^\mathrm{B} + R^\mathrm{C}) \frac{\D U_i}{\D t} + \left( 1 + \frac{R^\mathrm{B} + R^\mathrm{C}}{R_3} \right) U_i =
     \sum_j \frac{E_{ij}^\mathrm{E}}{C_1 R_{ij}^\mathrm{E}} \chi_{i 21} * \gamma_{ij}^\mathrm{E}  +
        \frac{R^\mathrm{D}}{C_1 (R^\mathrm{A} + R^\mathrm{D})} \chi_{i 21} * J_i  + \\
    \sum_k \frac{E_{ik}^\mathrm{I}}{C_2 R_{ik}^\mathrm{I}} \chi_{i 22} * \gamma_{ik}^\mathrm{I}  +
        \frac{R^\mathrm{D}}{C_2 (R^\mathrm{A} + R^\mathrm{D})} \chi_{i 22} * J_i + \frac{1}{C_2 (R^\mathrm{B} + R^\mathrm{C})} \chi_{i 22} * U_i \:.
\end{multline}
Utilizing the convolutions \pref{eq:GS:conv} once more, yields
\begin{multline}\label{eq:GS:nd2}
    C_3 (R^\mathrm{B} + R^\mathrm{C}) \frac{\D U_i}{\D t} + \left( 1 + \frac{R^\mathrm{B} + R^\mathrm{C}}{R_3} \right) U_i =
    \sum_j \frac{E_{ij}^\mathrm{E}}{C_1 R_{ij}^\mathrm{E}} \chi_{i 21} * \eta^\mathrm{E} * a_j +
    \sum_k \frac{E_{ik}^\mathrm{I}}{C_2 R_{ik}^\mathrm{I}} \chi_{i 22} * \eta^\mathrm{I} * a_k + \\
    \frac{R^\mathrm{D}}{C_1 (R^\mathrm{A} + R^\mathrm{D})} \chi_{i 21} * J_i  +
        \frac{R^\mathrm{D}}{C_2 (R^\mathrm{A} + R^\mathrm{D})} \chi_{i 22} * J_i + \frac{1}{C_2 (R^\mathrm{B} + R^\mathrm{C})}
        \chi_{i 22} * U_i \:.
\end{multline}
which becomes
\begin{multline}\label{eq:GS:nd3}
    C_3 (R^\mathrm{B} + R^\mathrm{C}) \frac{\D U_i}{\D t} + \left( 1 + \frac{R^\mathrm{B} + R^\mathrm{C}}{R_3} \right) U_i -
    \frac{1}{C_2 (R^\mathrm{B} + R^\mathrm{C})} \chi_{i 22} * U_i = \\
    \sum_j \frac{E_{ij}^\mathrm{E}}{C_1 R_{ij}^\mathrm{E}} \psi_{i 21} * a_j +
    \sum_k \frac{E_{ik}^\mathrm{I}}{C_2 R_{ik}^\mathrm{I}} \psi_{i 22} * a_k +
    \frac{R^\mathrm{D}}{R^\mathrm{A} + R^\mathrm{D}} \left[
        \frac{1}{C_1} \chi_{i 21}   +
        \frac{1}{C_2} \chi_{i 22} \right] * J_i
\end{multline}
after inserting the Green's functions \pref{eq:GS:green2a} and \pref{eq:GS:green2b} again. Equation \pref{eq:GS:nd3} together with \pref{eq:GS:srate}, \pref{eq:GS:thres} and \pref{eq:GS:reset} determine the dynamics of a network with three-compartment pyramidal neurons.


\section{Leaky Integrate-and-Fire Model}
\label{sec:GS:lif}

The most serious difficulty for dealing with the neurodynamical evolution equations (\ref{eq:GS:nd3}, \ref{eq:GS:srate}, \ref{eq:GS:thres}, \ref{eq:GS:reset}) and the DFP observation equation \pref{eq:GS:dfp} is the inhomogeneity of the matrix Green's function $\vec{X}_i(t, t')$ involved through the time-ordering operator and the time-dependence of $\vec{H}_i(t)$.


\subsection{Simplification}
\label{sec:GS:simp}

In a first approximation $\vec{H}_i$ becomes time-independent by neglecting the shunting terms \cite{BressloffTaylor93b, BressloffTaylor93a}. Then, the matrix Green's function $\vec{X}_i(t, t')$ becomes
\begin{equation}\label{eq:GS:homgreen}
    \vec{X}(t, t') = \vec{X}(t - t') = \E^{(t - t') \vec{H}} = \vec{Q}^{(t - t')}
\end{equation}
with
\begin{equation}\label{eq:GS:qmat}
    \vec{Q} = \E^{\vec{H}}
\end{equation}
and
\begin{equation}\label{eq:GS:hmat1}
    \vec{H} =
    \begin{pmatrix}
      \frac{1}{C_1} \left(
      -\frac{1}{R_1} + \frac{1}{R^\mathrm{A} + R^\mathrm{D}} \right)
      &
      -\frac{1}{C_1 (R^\mathrm{A} + R^\mathrm{D})} \\
      \frac{1}{C_2 (R^\mathrm{A} + R^\mathrm{D})}
      &
      \frac{1}{C_2}\left(
      -\frac{1}{R_2} - \frac{1}{R^\mathrm{A} + R^\mathrm{D}} - \frac{1}{R^\mathrm{B} + R^\mathrm{C}} \right)
    \end{pmatrix} \:,
\end{equation}
i.e. the transition matrix $\vec{H}$, and consequently also the Green's function, do not depend on the actual neuron index $i$ any more. In this case, analytical methods can be employed \cite{Bressloff94}.

However, for the present purpose, we employ another crucial simplification by choosing the electrotonic parameters in such a way that $\chi_{22}(t) \approx \delta(t)$. By virtue of this choice the dendritic filtering of compartment 2 is completely neglected. Then, the neural evolution equation \pref{eq:GS:nd3} turns into
\begin{multline}\label{eq:GS:nd4}
    C_3 (R^\mathrm{B} + R^\mathrm{C}) \frac{\D U_i}{\D t} + \left( 1 + \frac{R^\mathrm{B} + R^\mathrm{C}}{R_3} -
    \frac{\hat{\tau}}{C_2 (R^\mathrm{B} + R^\mathrm{C})} \right) U_i = \\
    \sum_j \frac{E_{ij}^\mathrm{E}}{C_1 R_{ij}^\mathrm{E}} \psi_{21} * a_j +
    \sum_k \frac{E_{ik}^\mathrm{I}}{C_2 R_{ik}^\mathrm{I}} \psi_{22} * a_k +
    \frac{R^\mathrm{D}}{R^\mathrm{A} + R^\mathrm{D}} \left[
        \frac{1}{C_1} \chi_{21}   +
        \frac{1}{C_2} \delta \right] * J_i \\
    C_3 (R^\mathrm{B} + R^\mathrm{C}) \frac{\D U_i}{\D t} +
    \frac{C_2 R_3 (R^\mathrm{B} + R^\mathrm{C}) + C_2 (R^\mathrm{B} + R^\mathrm{C})^2 - \hat{\tau} R_3}{C_2 R_3 (R^\mathrm{B} + R^\mathrm{C})} U_i = \\
    \sum_j \frac{E_{ij}^\mathrm{E}}{C_1 R_{ij}^\mathrm{E}} \psi_{21} * a_j +
    \sum_k \frac{E_{ik}^\mathrm{I}}{C_2 R_{ik}^\mathrm{I}} \psi_{22} * a_k +
    \frac{R^\mathrm{D}}{R^\mathrm{A} + R^\mathrm{D}} \left[
        \frac{1}{C_1} \chi_{21}   +
        \frac{1}{C_2} \delta \right] * J_i
\end{multline}
where all kernels lost their neuron index $i$. Additionally, some time constant $\hat{\tau}$ results from the temporal convolution. Multiplying next with
\begin{equation}\label{eq:GS:propr}
    r = \frac{C_2 R_3 (R^\mathrm{B} + R^\mathrm{C})}{C_2 R_3 (R^\mathrm{B} + R^\mathrm{C}) + C_2 (R^\mathrm{B} + R^\mathrm{C})^2 - \hat{\tau} R_3}
\end{equation}
yields a leaky integrate-and-fire (LIF) model
\begin{equation}\label{eq:GS:lif}
    \tau \frac{\D U_i}{\D t} + U_i = \sum_j w_{ij}^\mathrm{E} \, \psi_{21} * a_j +
    \sum_k w_{ik}^\mathrm{I} \, \psi_{22} * a_k +
    \kappa \left[
        \frac{1}{C_1} \chi_{21}   +
        \frac{1}{C_2} \delta \right] * J_i
\end{equation}
where we have introduced the following parameters:
\begin{itemize}
    \item \emph{time constant}
    \begin{equation}
    \label{eq:timeconst}
      \tau = r C_3 (R^\mathrm{B} + R^\mathrm{C})
    \end{equation}
    \item \emph{excitatory synaptic weights}
    \begin{equation}
    \label{eq:pexweig}
      w_{ij}^\mathrm{E} = r \frac{E_{ij}^\mathrm{E}}{C_1 R_{ij}^\mathrm{E}} > 0
    \end{equation}
    \item \emph{inhibitory synaptic weights}
    \begin{equation}
    \label{eq:pinweig}
      w_{ik}^\mathrm{I} = r \frac{E_{ik}^\mathrm{I}}{C_2 R_{ik}^\mathrm{I}} < 0
    \end{equation}
    \item \emph{diffusion coefficient}
    \begin{equation}
    \label{eq:somdifres}
      \kappa  = r \frac{R^\mathrm{D}}{R^\mathrm{A} + R^\mathrm{D}}  \:.
    \end{equation}
\end{itemize}

Moreover, we make the same approximation for the DFP \pref{eq:GS:dfp} and obtain
\begin{multline}\label{eq:GS:dfp1}
  \Phi_i =
  \frac{R^\mathrm{D}}{R^\mathrm{A} + R^\mathrm{D}}
    \Bigg\{
    \sum_j \frac{E_{ij}^\mathrm{E}}{C_1 R_{ij}^\mathrm{E}} (\psi_{11} - \psi_{21}) * a_j  +
    \sum_k \frac{E_{ik}^\mathrm{I}}{C_2 R_{ik}^\mathrm{I}} (\psi_{12} - \delta) * a_k + \\
        \Bigg[ \frac{R^\mathrm{D}}{C_1 (R^\mathrm{A} + R^\mathrm{D})} (\chi_{11} - \chi_{21}) +
                  \frac{R^\mathrm{D}}{C_2 (R^\mathrm{A} + R^\mathrm{D})} (\chi_{12} - \delta) - R^\mathrm{A} \delta
        \Bigg]* J_i  + \\
        \frac{1}{C_2 (R^\mathrm{B} + R^\mathrm{C})} (\chi_{12} - \delta) * U_i  \Bigg\}
\end{multline}
as the observation equation of the LIF model.

Equations (\ref{eq:GS:lif}, \ref{eq:GS:dfp1}) still exhibit some redundancy, seeing that the kernel $\psi_{21}$ always relates to excitatory synapses while the kernel $\psi_{22}$ refers to inhibitory synapses. We could thus absorb these kernel indices into the presynaptic neuron indices by introducing new kernels
\begin{eqnarray}
    \label{eq:GS:kern1}
        \psi_j &=& \begin{cases}
                        \psi_{21} & \quad:\quad j \text{ excitatory} \\
                        \psi_{22} & \quad:\quad j \text{ inhibitory}
                    \end{cases} \\
    \label{eq:GS:kern2}
        \zeta_j &=& \begin{cases}
                        \psi_{11} - \psi_{21} & \quad:\quad j \text{ excitatory} \\
                        \psi_{12} - \delta       & \quad:\quad j \text{ inhibitory} \:.
                    \end{cases}
\end{eqnarray}
These kernels entail
\begin{eqnarray}
    \label{eq:GS:lif2}
        \tau \frac{\D U_i}{\D t} + U_i &=& \sum_j w_{ij} \, \psi_j * a_j +
        \kappa \left[ \frac{1}{C_1} \chi_{21} + \frac{1}{C_2} \delta \right] * J_i        \\
    \label{eq:GS:dfp2}
      \Phi_i &=&
      \frac{R^\mathrm{D}}{r(R^\mathrm{A} + R^\mathrm{D})}
        \Bigg\{
        \sum_j w_{ij} \, \zeta_j * a_j  + \nonumber \\
&&
            \Bigg[ \frac{R^\mathrm{D}}{C_1 (R^\mathrm{A} + R^\mathrm{D})} (\chi_{11} - \chi_{21}) +
                      \frac{R^\mathrm{D}}{C_2 (R^\mathrm{A} + R^\mathrm{D})} (\chi_{12} - \delta) - R^\mathrm{A} \delta
            \Bigg]* J_i  + \nonumber\\
&&
            \frac{1}{C_2 (R^\mathrm{B} + R^\mathrm{C})} (\chi_{12} - \delta) * U_i  \Bigg\}
\end{eqnarray}
after also dropping the redundant excitatory/inhibitory superscripts. Thus, the indices $i, j$  now extend over the entire network of $K$ units.

Because the relevance of diffusion currents is controversially discussed in the literature \cite{GabrielEA96a, GabrielEA96b, GabrielEA96c, BedardDestexhe09, BedardRodriguesEA10}, we neglect these for further simplification: $J_i = 0$ which leads to
\begin{eqnarray}
    \label{eq:GS:lif3}
        \tau \frac{\D U_i}{\D t} + U_i &=& \sum_j w_{ij} \, \psi_j * a_j  \\
    \label{eq:GS:dfp3}
      \Phi_i &=&
      \frac{R^\mathrm{D}}{r(R^\mathrm{A} + R^\mathrm{D})}
        \Bigg\{ \sum_j w_{ij} \, \zeta_j * a_j  +
       \frac{1}{C_2 (R^\mathrm{B} + R^\mathrm{C})} (\chi_{12} - \delta) * U_i  \Bigg\} \:. \nonumber \\
      && {}
\end{eqnarray}


\subsection{Simulation}
\label{sec:GS:sim}

We have extensively discussed the system (\ref{eq:GS:lif3}, \ref{eq:GS:dfp3}) under the further assumption $\chi_{12} = \delta$ in a recent paper \cite{GrabenRodrigues13a}, and herein we present further numerical simulations under different external stimulation input. In particular, we simulate a cortical tissue as a LIF network~(\ref{eq:GS:lif3}), comprising of 1000 interneurons and 4000 pyramidal neurons interconnected randomly via an Erd\H{o}s-R\'{e}nyi graph with connection probability of 0.2. We refer the reader to \cite{GrabenRodrigues13a} on how the somatic, dendritic and extracellular electrotonic parameters are estimated and how these are related to the phenomenological parameters of Mazzoni et. al. \cite{MazzoniPanzeriEA08}. All other parameters such as steady state voltages, refractory period, synaptic latencies, thresholds and others can also be found therein. Thalamic inputs are the only source of noise, which attempt to account for both cortical heterogeneity and spontaneous activity. This is achieved by modeling a two level noise, where the first level is an Ornstein-Uhlenbeck process superimposed with a constant or periodic signal and the second level is a time varying inhomogeneous Poisson process. Thus, we have the following time varying rate, $\lambda(t)$, that feeds into inhomogeneous Poisson process:
\begin{eqnarray}
    \tau_n \frac{\D n(t)}{\D t} &=& - n(t) + \sigma_n \sqrt{\frac{2}{\tau_n}} \, \eta(t) \\
    \lambda(t) &=& [c_0 + n(t)]_{+}
\end{eqnarray}
where $\eta(t)$ represents Gaussian white noise, $c_0$ represents a constant signal (but equally could be periodic or other), and the operation $[\cdot]_{+}$ is the threshold-linear function, $[x]_{+}=x$ if $x > 0$, $[x]_{+}=0$ otherwise, which circumvents negative rates. The constant signal $c_0$ can range between $1.2$ to $2.6$ spikes/ms. Note also that periodic or more complex `naturalistic' signals can be applied, but we have herein kept it simple just for illustrative purposes. The parameters of the Ornstein-Uhlenbeck process are $\tau_n = 16~\mathrm{ms}$ and the standard deviation $\sigma_n = 0.4$ spikes/ms, also refer to \cite{MazzoniPanzeriEA08} for these parameters.

\begin{figure}[H]
 \centering
  \includegraphics[width=0.9\textwidth]{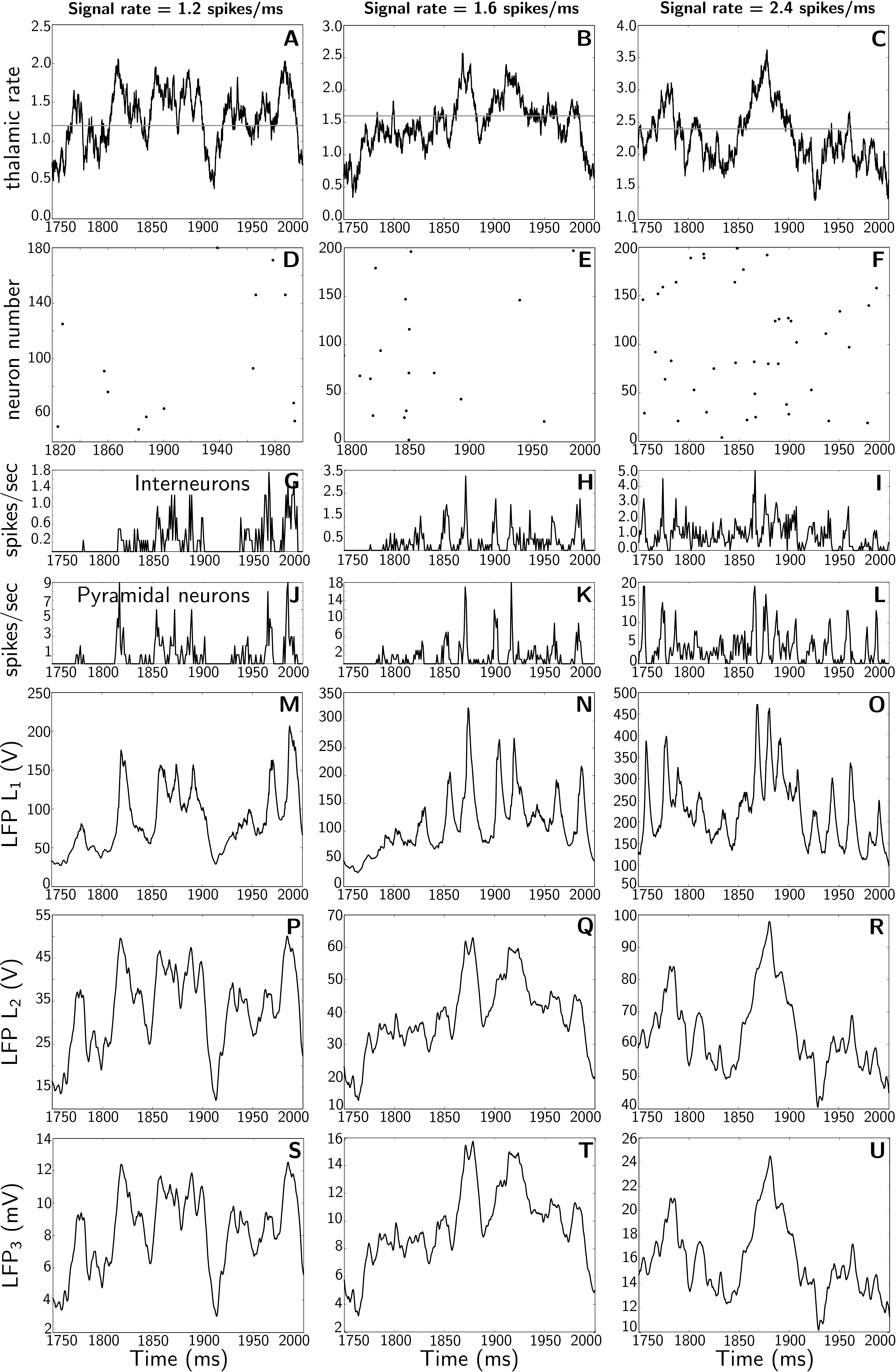}
\caption{\label{fig:lfp1} Dynamics of the network and LFP comparisons: The three columns represent different runs of the network for three different rates, 1.2, 1.6 and 2.4 spikes/ms. In each column, all panels show the same 250 ms (extracted from 2 seconds simulations). The top panels (A-C) represent thalamic inputs with the different rates. The second top panels (D-F) correspond to a raster plot of the activity of 200 pyramidal neurons. Panels (G-I) depict average instantaneous firing rate (computed on a 1ms bin) of interneurons and panels (J-L) correspond to average instantaneous firing rate of pyramidal neurons. Panels (M-O) show the LFP $L_1$ (\Eq{eq:lfp1}) which is the Mazzoni et. al measure~\cite{MazzoniPanzeriEA08}. Panels (P-R) and (S-U) depict our proposed LFP measures $L_2$ (\Eq{eq:lfp2}) and $L_3$ (\Eq{eq:lfp3}), respectively. Note and compare the different order of magnitudes between the three LFP measures.}
\end{figure}

The network simulations were run under the \emph{Brian Simulator}, which is a Python based environment~\cite{goodman2009brian}. We focus on resistive extracellular case and compare between our DFP $\Phi_i$ measure~(\ref{eq:GS:dfp3}) and the Mazzoni LFP measure (MPLB) defined herein as the sum of the moduli of excitatory and inhibitory synaptic currents:
\begin{eqnarray}
 V_i^\mathrm{MPLB}  = \sum_j |w_{ij}^\mathrm{E} \, \psi_{21} * a_j| + \sum_k |w_{ik}^\mathrm{I} \, \psi_{22} * a_k | \label{MLFP}
\end{eqnarray}
In addition, Mazzoni et. al.~\cite{MazzoniPanzeriEA08} take the sum of $V_i^\mathrm{MPLB}$ across all pyramidal neurons. To provide a comparison we will also consider the sum of our proposed DFP measure~(\ref{eq:GS:dfp3}), but also contrast it with its average. Thus we compare the following models of local field potentials:
\begin{eqnarray}
    \label{eq:lfp1} L_1 &=& \sum_i V_i^\mathrm{MPLB} \\
    \label{eq:lfp2} L_2 &=& \sum_i \Phi_i \\
    \label{eq:lfp3} L_3 &=& \frac{1}{P} \sum_i \Phi_i  \:,
\end{eqnarray}
where $P$ is the number of pyramidal neurons. Subsequently, we run the network for two seconds with three different noise levels, specifically, receiving a constant signals with rates $1.2$, $1.6$ and $2.4$ spikes/ms as depicted in Fig.~\ref{fig:lfp1}. We report two main striking differences between LFP measures \Eq{eq:lfp1}, \Eq{eq:lfp2} and \Eq{eq:lfp3}, namely in frequency and in amplitude. The $L_1$ responds instantaneously to the spiking network activity by means of high frequency oscillations. Moreover, $L_1$ exhibits a large amplitude overestimating experimental LFP/EEG measurements that typically vary between 0.5 to 2 mV \cite{SchomerSilva11, lakatos2005oscillatory}. In contrast, $L_2$ and $L_3$ respond more smoothly to population activity and it is noticeable that our LFP estimates represent the low-pass filtered thalamic input. Clearly, both $L_2$ and $L_3$ have same time profile, however, the $L_3$ measures comparably with experimental LFP, that is, in the order of millivolts (although its not contained within the experimental range 0.5 to 2 mV). Thus we do concede that further work is required to improve our estimates. A minor improvement could be attained by applying a weighted average to mimic the distance of an electrode to a particular neuron. However, more biophysical modeling work is in demand as other neural effects, such as neuron-glia interaction and other effects, could be required to bring down these estimates to experimental values.

\section{Continuum Model}
\label{sec:GS:conti}

So far we have considered the electrical properties of neural networks containing cortical pyramidal cells by means of equivalent circuits of a three-compartment model. In order to link these properties to the electromagnetic field in extracellular space, we need an embedding of the network topology into physical metric space $\mathbb{R}^3$. This is most easily achieved in the continuum limit of neural field theory.


\subsection{Rate Model}
\label{sec:GS:rate}

Starting with the approximation from \Sec{sec:GS:lif}, we first transform our LIF approach into a rate model. According to \Eq{eq:GS:srate}, a spike train is represented by a sum over delta functions. In order to obtain the number of spikes in a time interval $[0, t]$, one has to integrate \Eq{eq:GS:srate}, yielding
\[
    n(t) = \int_0^t a(t') \, \D t' \:.
\]
Then, the instantaneous spike rate per unit time is formally regained as the original signal \Eq{eq:GS:srate}, through
\begin{equation}\label{eq:GS:spikerate}
    \frac{\D}{\D t} n(t) = a(t)  \:.
\end{equation}

A spike train $a(t)$ arriving at the presynaptic terminal of an axon causes changes in the conductivity of voltage-gated calcium channels. Therefore, calcium current flows into the synaptic button evoking the release of neurotransmitter into the synaptic cleft which is essentially a stochastic Bernoulli process \cite{Graben08a, KandelSchwartzEA91}, where the probability $P(k)$ for releasing $k$ transmitter vesicles obeys a binomial distribution
\begin{equation}\label{eq:GS:binom}
    P(k) = {Y \choose k} \, p^k (1 - p)^{Y - k} \:,
\end{equation}
with $Y$ the number of allocated vesicles in the button and $p$ the elementary probability that an arriving action potential releases one vesicle.

In the limit of large numbers, the binomial distribution can be replaced by a normal distribution
\begin{equation}\label{eq:GS:normal}
    P(k) = \frac{1}{\sqrt{2 \pi y (1 - p)}} \, \exp{\left[-\frac{(k - y)^2}{2 y (1 - p)} \right]} \:,
\end{equation}
where $y = Y p$ is the average number of allocated transmitter vesicles. Due to this stochasticity of synaptic transmission, even the dynamics of a single neuron should be treated in terms of statistic ensembles in probability theory. Hence, we describe the state variables $U_i(t)$ by a normal distribution density $\rho(u, t)$ with mean $\bar{U}(t)$ and variance $\sigma^2$, and determine the firing probability as
\begin{equation}\label{eq:GS:probfeuer}
    r(t) = \Pr(U(t) \ge \theta) = \int_\theta^\infty \rho(u, t) \, \D u
    = \frac{1}{2}\, \erfc \left(\frac{\theta - \bar{U}}{\sqrt{2} \sigma}\right)  \:,
\end{equation}
with `erfc' as the complementary error function accounting for the cumulative probability distribution. Thereby, the stochastic threshold dynamics is characterized by the typical sigmoidal activation functions. In computational neuroscience, the complementary error function is often approximated by the logistic function
\[
    f(u) =  \frac{1}{1 +  \E^{-\gamma (u - \theta)}}
\]
with parameters gain $\gamma$ and threshold $\theta$. Using $f$ as nonlinear activation function, the firing probability $r(t) = f(U(t))$ for mean membrane potential $U$ is closely related to the instantaneous spike rate $a(t)$ [\Eq{eq:GS:spikerate}] via
\begin{equation}\label{eq:GS:firerat}
    a(t) = a_{\max} \, r(t) = a_{\max} \, f(U)
\end{equation}
with maximal spike rate $a_{\max}$ which can be absorbed by $f$:
\begin{equation}\label{eq:GS:logi}
    f(u) =  \frac{a_{\max}}{1 +  \E^{-\gamma (u - \theta)}} \:.
\end{equation}

Inserting \pref{eq:GS:firerat} and \pref{eq:GS:logi} into our LIF model \pref{eq:GS:lif2} yields a leaky integrator rate (LIR) model \cite{BressloffTaylor91, GrabenLiebscherKurths08}
\begin{equation}\label{eq:GS:lir}
        \tau \frac{\D U_i}{\D t} + U_i = \sum_j w_{ij} \, \psi_j * f(U_j) +
        \kappa \left[ \frac{1}{C_1} \chi_{21} + \frac{1}{C_2} \delta \right] * J_i  \:.
\end{equation}


\subsection{Neuroelectrodynamics}
\label{sec:GS:ned}

Next we perform the continuum limit $U_i(t) \to u_n(\vec{x}, t)$ \emph{\`{a} la physique} under the assumption of identical neural properties within each population. The sum in \Eq{eq:GS:lir} may converge under suitable smoothness assumptions upon the synaptic weight matrix $w_{ij}$ and the convolution kernels.\footnote{
    A rigorous treatment of the continuum limit for neural networks requires techniques from stochastic analysis such as mean-field approaches or system-size expansions as carried out by Faugeras, Touboul and Cessac \cite{FaugerasTouboulCessac08} and Bressloff \cite{Bressloff10} (see also the chapter of Bressloff in this volume).
}
Then a continuous two-dimensional vector $\vec{x} \in \Gamma_n$ replaces the neuron index $i$, while $n$ becomes a population index. The population layers $\Gamma_n$ become two-dimensional manifolds embedded in three-dimensional physical space such that $\vec{x} \in \Gamma_n$ is a two-dimensional projection of a vector $\vec{r} \in C \subset \mathbb{R}^3$ ($C$ denoting cortex). Or, likewise, $\vec{r} = (\vec{x}, z)$, as indicated in \Fig{fig:GS:pyramid}.

As a result, \Eq{eq:GS:lir} passes into the Amari equation \cite{Amari77b}
\begin{equation}\label{eq:GS:nft}
    \tau \frac{\partial}{\partial t} u_i(\vec{x}, t) + u_i(\vec{x}, t) = \sum_k \int_{\Gamma_k}
         w_{ik}(\vec{x}, \vec{x}') \, \psi(\vec{x}', t) * f(u_k(\vec{x}', t)) \, \D\vec{x}' + h_i(\vec{x}, t)
\end{equation}
with input current $h_i(\vec{x}, t)$ delivered to neuron layer $i$ at site $\vec{x} \in \Gamma_i$. The synaptic weight kernels $w_{ik}(x, x')$ and the synaptic-dendritic impulse response $\psi(\vec{x}', t)$ are obtained from the synaptic weight matrix, and from the Green's functions $\psi_j(t)$, respectively.

This neural field equation is complemented by the continuum limit of the extracellular dendritic dipole current density through cross section area $A$ with normal vector $\vec{n}_A$, shown in \Fig{fig:GS:pyramid}, which is obtained from \pref{eq:GS:dendcurrent3}, i.e.
\begin{multline}\label{eq:GS:dc}
    \vec{j}(\vec{r}, t) = \lim_{i \to \vec{x}} \frac{\vec{n}_A}{A} I_{i1} =  \\
         = \sum_k \int_{\Gamma_k} \tilde{w}_{1k}(\vec{r}, \vec{r}') \, \zeta(\vec{r}', t) * f(u_k(\vec{r}', t)) + \xi_1(t) * \vec{j}^\mathrm{D}(\vec{r}, t) +  \xi_2(t) * u_1(\vec{r}, t) \:,
\end{multline}
where we have introduced a modified synaptic weight kernel $\tilde{w}$ and two new convolution kernels $\xi_j$ that are related to the electrotonic parameters of the discrete model \pref{eq:GS:dendcurrent3}. The proper diffusion current $\vec{j}^\mathrm{D}(\vec{r}, t)$ must be related to the gradient of the charge density $\rho(\vec{r},t)$ according to Fick's law
\begin{equation}\label{eq:GS:fick}
     \vec{j}^\mathrm{D}(\vec{r}, t) = - \vec{D}(\vec{r}, t) \nabla \rho(\vec{r}, t) \:,
\end{equation}
where the diffusion tensor $\vec{D}(\vec{r}, t)$ accounts for inhomogeneities and unisotropies of extracellular fluid, as being reflected by \emph{diffusion tensor imaging} (DTI) \cite{BedardKroegerDestexhe04, BedardDestexhe12}. For layer 1 of pyramidal neurons the input is then given by the diffusion current $\vec{j}^\mathrm{D}(\vec{r}, t)$. Therefore, the input to the Amari equation \pref{eq:GS:nft} becomes
\begin{equation}\label{eq:GS:difinp}
    h_i(\vec{x}, t) = - \delta_{i1} \kappa A \, \vec{D}(\vec{r}, t) \cdot \nabla \rho(\vec{r}, t)
\end{equation}
with Kronecker's $\delta_{ik}$ and appropriately redefined $\kappa$.

For further treatment of the electrodynamics of neural fields in linear but inhomogeneous and unisotropic media, we need Ohm's law
\begin{equation}\label{eq:GS:ohm}
     \vec{j}^\mathrm{\Omega}(\vec{r}, t) = \vec{\sigma}(\vec{r}, t) \cdot \vec{E}(\vec{r}, t) \:,
\end{equation}
where $\vec{\sigma}(\vec{r}, t)$ is the conductivity tensor and $\vec{E}$ the electric field strength. In case of negligible magnetic fields, we can introduce the dendritic field potential $\varphi$ via
\begin{equation}\label{eq:GS:pot}
     \vec{E} = - \nabla \varphi \:.
\end{equation}

Diffusion current \pref{eq:GS:fick} and Ohmic current \pref{eq:GS:ohm} together entail the Nernst-Planck equation \cite{JohnstonWu97, Tuckwell88a}
\begin{equation}
 \label{eq:GS:npe}
 \vec{j} = - \vec{D} \cdot \nabla \rho + \vec{\sigma} \cdot \vec{E} \:.
\end{equation}

In the diffusive and conductive extracellular fluid, we additionally have
\begin{itemize}
    \item \emph{Einstein's relation} \cite{Einstein06}
    \begin{equation}
    \label{eq:GS:einstein}
        \vec{D} = k_\mathrm{B} T q \vec{\mu}
    \end{equation}
    \item \emph{conductivity}
    \begin{equation}
    \label{eq:GS:conduct}
        \vec{\sigma} = \vec{\mu} \rho \:,
    \end{equation}
\end{itemize}
where $k_\mathrm{B}$ denotes the thermodynamic Boltzmann constant, $T$ temperature, $q$ the ion charge, and
$\vec{\mu}$ the ion's mobility tensor related to the fluid's viscosity \cite{JohnstonWu97, Tuckwell88a}. Inserting (\ref{eq:GS:einstein}, \ref{eq:GS:conduct}) into \pref{eq:GS:npe} yields
\begin{equation}\label{eq:GS:npe2}
 \vec{j} = - k_\mathrm{B} T q \vec{\mu} \cdot \nabla \rho + \vec{\mu} \cdot \vec{E} \rho \:.
\end{equation}

This form of the Nernst-Planck equation has to be augmented by a continuity equation
\begin{equation}
 \label{eq:GS:conti}
 \nabla \cdot \vec{j} + \frac{\partial \rho}{\partial t} = 0
\end{equation}
reflecting the conservation of charge as a consequence of Maxwell's equations, and the by first Maxwell equation
\begin{equation}
 \label{eq:GS:max1}
 \nabla \cdot \vec{D} = \rho
\end{equation}
where
\begin{equation}
 \label{eq:GS:permitt}
 \vec{D} = \vec{\epsilon} \cdot \vec{E}
\end{equation}
introduces the electrical permittivity tensor $\vec{\epsilon}$.

Inserting \pref{eq:GS:permitt} into \pref{eq:GS:max1} first gives
\begin{equation}
 \label{eq:GS:max11}
 \vec{\epsilon} \cdot (\nabla \cdot \vec{E}) = \rho - (\nabla \cdot \vec{\epsilon}) \cdot \vec{E} \:.
\end{equation}
Next, we take the divergence of the Nernst-Planck equation \pref{eq:GS:npe2}, which yields after consideration of the continuity equation \pref{eq:GS:conti}
\begin{eqnarray*}
    \nabla \cdot \vec{j} &=& - k_\mathrm{B} T q \nabla \cdot (\vec{\mu} \cdot \nabla \rho)  + \nabla \cdot (\vec{\mu} \cdot \vec{E} \rho) \\
    - \vec{\epsilon} \frac{\partial \rho}{\partial t} &=& - k_\mathrm{B} T q \vec{\epsilon} \cdot
        [ (\nabla \cdot \vec{\mu}) \cdot (\nabla \rho) + \vec{\mu} \Delta \rho ] +
                \vec{\epsilon} \cdot (\nabla \cdot \vec{\mu}) \cdot \vec{E} \rho + \vec{\epsilon} \cdot \vec{\mu} \cdot (\nabla \cdot \vec{E}) \rho + \vec{\epsilon} \cdot \vec{\mu} \cdot \vec{E} \cdot \nabla \rho \:.
\end{eqnarray*}

Introducing the commutator $[\vec{\epsilon} , \vec{\mu}] = \vec{\epsilon} \cdot \vec{\mu}  - \vec{\mu} \cdot \vec{\epsilon}$, we can write
\begin{eqnarray*}
    - \vec{\epsilon} \frac{\partial \rho}{\partial t} &=& - k_\mathrm{B} T q \vec{\epsilon} \cdot
        [ (\nabla \cdot \vec{\mu}) \cdot (\nabla \rho) + \vec{\mu} \Delta \rho ] +
                \vec{\epsilon} \cdot (\nabla \cdot \vec{\mu}) \cdot \vec{E} \rho + \\
&&
                [\vec{\epsilon} , \vec{\mu}] \cdot (\nabla \cdot \vec{E}) \rho + \vec{\mu} \cdot \rho^2 - \vec{\mu} \cdot (\nabla \cdot \vec{\epsilon}) \cdot \vec{E} \rho +
                \vec{\epsilon} \cdot \vec{\mu} \cdot \vec{E} \cdot \nabla \rho \:,
\end{eqnarray*}
where we have also utilized \pref{eq:GS:max11}.

Using the Nernst-Planck equation \pref{eq:GS:npe2} once more, we eliminate the electric field
\begin{equation}\label{eq:GS:npe3}
    \vec{E} =  \vec{\mu}^{-1} \cdot \left( \frac{\vec{j} + k_\mathrm{B} T q \vec{\mu} \cdot \nabla \rho}{\rho} \right)
\end{equation}
thus
\begin{multline}\label{eq:GS:rhopde}
    - \vec{\epsilon} \frac{\partial \rho}{\partial t} = - k_\mathrm{B} T q \vec{\epsilon} \cdot
        [ (\nabla \cdot \vec{\mu}) \cdot (\nabla \rho) + \vec{\mu} \Delta \rho ] +
                \vec{\epsilon} \cdot \nabla (\ln \vec{\mu}) \cdot ( \vec{j} + k_\mathrm{B} T q \vec{\mu} \cdot \nabla \rho ) + \\
                [\vec{\epsilon} , \vec{\mu}] \cdot \nabla \cdot \left\{ \vec{\mu}^{-1} \cdot \left( \frac{\vec{j} + k_\mathrm{B} T q \vec{\mu} \cdot \nabla \rho}{\rho} \right) \right\} \rho + \vec{\mu} \cdot \rho^2 - \vec{\mu} \cdot (\nabla \cdot \vec{\epsilon}) \cdot \vec{\mu}^{-1} \cdot
                ( \vec{j} + k_\mathrm{B} T q \vec{\mu} \cdot \nabla \rho ) + \\
                \vec{\epsilon} \cdot \left( \frac{\vec{j} + k_\mathrm{B} T q \vec{\mu} \cdot \nabla \rho}{\rho} \right) \cdot \nabla \rho \:.
\end{multline}

The solution of \pref{eq:GS:rhopde} provides the extracellular charge density $\rho(\vec{x}, t)$ dependent on the extracellular driving currents $\vec{j}$ that are delivered by the neural field equations (\ref{eq:GS:nft}, \ref{eq:GS:dc}, \ref{eq:GS:difinp}). Inserting both $\rho(\vec{x}, t)$ and $\vec{j}$ into \pref{eq:GS:npe3} yields the DFP
\begin{equation}\label{eq:GS:npe4}
    - \nabla \varphi  =  \vec{\mu}^{-1} \cdot \left( \frac{\vec{j} + k_\mathrm{B} T q \vec{\mu} \cdot \nabla \rho}{\rho} \right)
\end{equation}
via \pref{eq:GS:pot}.

These equations of neuroelectrodynamics can be considerably simplified by assuming a homogeneous and isotropic medium. In that case \pref{eq:GS:rhopde} reduces to
\begin{equation}\label{eq:GS:rhosimp}
        - \vec{\epsilon} \frac{\partial \rho}{\partial t} = - k_\mathrm{B} T q \vec{\epsilon} \cdot
        \vec{\mu} \Delta \rho  +
                \vec{\mu} \cdot \rho^2 +
                \vec{\epsilon} \cdot \left( \frac{\vec{j} + k_\mathrm{B} T q \vec{\mu} \cdot \nabla \rho}{\rho} \right) \cdot \nabla \rho \:,
\end{equation}
which is a kind of Fokker-Planck equation for the charge density. Taking only the first term of the r.h.s. into account, we obtain a diffusion equation whose stationary solution gives rise to the Warburg impedance of extracellular space \cite{BedardDestexhe09, BedardRodriguesEA10, BedardDestexhe12, SkaleDolecekSlemnik07}.


\section{Discussion}
\label{sec:GS:disc}
In this contribution we outlined a biophysical theory for the coupling of microscopic neural activity to the electromagnetic field as described by the Maxwell equations, in order to adequately explain neural field potentials, such as DFP, LFP, and EEG. To that aim we have started from the widely accepted assumption, that cortical LFP/EEG mostly reflect extracellular dipole currents of pyramidal cells \cite{NunezSrinivasan06, SchomerSilva11}. This assumption has lead us to recent work suggesting that both Ohmic and diffusion currents contribute to LFP/EEG generation \cite{GabrielEA96a, GabrielEA96b, GabrielEA96c, BedardDestexhe09, BedardRodriguesEA10}. In addition, the assumption has placed a further challenge in that the geometry of the cortical pyramidal cells should be incorporated. Accounting for the geometry of the cell seemed to imply that one loses the computational efficiency of point models and we had to resort to compartmental models. However, herein we have proposed a framework showing how to circumvent these apparent difficulties to finally derive a biophysically plausible observation model for the Amari neural field equation\cite{Amari77b}, with additional dipole currents coupled to the Maxwell's equations.

We have first proposed a full-fledged three-compartment model of a single pyramidal cell decomposed into the apical dendritic tree for the main of excitatory synapses, the soma and the perisomatic dendritic tree that harbors mainly the inhibitory synapses, and the axon hillock exhibiting the neural spiking mechanism. In addition, the extracellular space was represented by incorporating both Ohmic and diffusive impedances, thus assuming that the total current through the extracellular fluid is governed by the Nernst-Planck equation. This has enabled us to account for the Warburg impedance. From this starting point and successive simplifications we have derived the evolution law of the circuit, represented as an integro-differential equation. In the continuum limit this evolution law went into the Amari neural field equation, augmented by an observation equation for dendritic dipole currents, that are coupled to Maxwell's equations for the electromagnetic field in extracellular fluid.

Moreover we have demonstrated how to simplify and derive from our proposed three-compartment model a standard LIF network which then have enabled us to compare our LFP measure with other LFP measures found in the literature. Herein, we specifically have chosen to compare with the Mazzoni et. al. work~\cite{MazzoniPanzeriEA08}, that proposed the LFP to be the sum of the moduli of inhibitory and excitatory currents. Thus, we have proceeded by mapping our biophysical electrotonic parameters to the phenomenological parameters implemented in Mazzoni's LFP model~\cite{MazzoniPanzeriEA08}. However, now with the advantage that our LFP measure accounts for the extracellular currents and the geometry of the cell. Subsequently, we have compared different simulation runs between our LIF network model and that of Mazzoni et al. \cite{MazzoniPanzeriEA08}. This comparison indicates that the Mazzoni et al. model systematically overestimates LFP amplitude by almost one order of magnitude and also systematically overestimates LFP frequencies. For more detailed discussion we refer the interested reader to \cite{GrabenRodrigues13a}.

At the present stage, we note that there is still a long way to fully explain the spatiotemporal characteristics of LFP and EEG. For example, the polarity reversals observed in experimental LFP/EEG as an electrode traverses different cortical layers are not accounted for in our current model. This is explained with the direction of the dipole currents, which is constrained, in the sense that current sources are situated at the perisomatic and current sinks are situated at apical dendritic tree. Taking this polarity as positive also entails positive DFP and LFP that could only change in strength. However, it is straightforward to adapt our model by fully incorporating cortical layers III and VI, for example. Yet another aspect that was not looked in the present work, was that of \emph{ephaptic interactions} \cite{FrohlichMcCormick10, HoltKoch99a, Jefferys95, RichardsonSchiffGluckman05} between neurons and the LFP which could act via a \emph{mean-field coupling} as an \emph{order parameter} thereby entraining the local populations to synchronized activity. A possible biophysical basis for this phenomena could be polarization of neurons induced by electric fields that are generated by ionic charges. As a consequence, this could alter the voltage dependent conductances, triggering changes of the neuronal dynamics, such as spiking and the activity of glia cells. We have not accounted for this effect in a biophysical sense yet, however, we could phenomenologically describe this mean-field coupling through a modulation of firing thresholds as outlined in \cite{Graben08a}.


\begin{acknowledgement}
We thank Axel Hutt, Jamie Sleigh, Viktor Jirsa and Dimitris Pinotsis for helpful comments improving this chapter. This research was supported by a DFG Heisenberg fellowship awarded to PbG (GR 3711/2-1).
\end{acknowledgement}


\end{document}